\documentclass[twocolumn,preprintnumbers,amsmath,amssymb]{revtex4}
\usepackage{graphicx}
\usepackage{dcolumn}
\usepackage{bm}
\usepackage[final]{pdfpages}
\begin{document}
\title{Non-monotonic anisotropy in charge conduction induced by the antiferrodistortive transition in metallic SrTiO$_{3}$}
\author{Qian Tao$^{1,2}$, Bastien Loret $^{1}$, Bin Xu $^{3,4}$,  Xiaojun Yang$^{1,2}$, Carl Willem Rischau$^{1}$, Xiao Lin$^{1}$,  Beno\^{\i}t Fauqu\'e$^{1}$  Matthieu J. Verstraete $^{3}$ and Kamran Behnia$^{1}$\email{kamran.behnia@espci.fr}}
\affiliation{(1) Laboratoire Physique et Etude de Mat\'{e}riaux-CNRS/ESPCI/UPMC \\ Paris, F-75005, France\\
(2) Department of Physics, Zhejiang University, Hangzhou 310027, China\\
(3) CESAM and European Theoretical Spectroscopy Facility, Universit\'{e} de Li\`{e}ge, B-4000 Li\`{e}ge, Belgium\\
(4) Physics Dept and Institute for Nanoscience and Engineering, Univ of Arkansas, Fayetteville, AR 72701, USA}
\date{June 17, 2016}

\begin{abstract}
Cubic SrTiO$_{3}$ becomes tetragonal below 105 K. The antiferrodistortive (AFD) distortion leads to clockwise and counter-clockwise  rotation of adjacent TiO$_{6}$ octahedra. This insulator becomes a metal upon the introduction of extremely low concentration of n-type dopants. However, signatures of the structural phase transition in charge conduction have remained elusive. Employing the Montgomery technique, we succeed in resolving the anisotropy of charge conductivity induced by the AFD transition, in the presence of different types of dopants. We find that the slight lattice distortion ($<6 \times 10^{-4}$) gives rise to a twenty percent anisotropy in charge conductivity, in agreement with the expectations of band calculations. Application of uniaxial strain amplifies the detectable anisotropy by  disfavoring one of the three possible tetragonal domains. In contrast with all other known anisotropic Fermi liquids, the anisotropy has opposite signs for elastic and inelastic scattering. Increasing the concentration of dopants leads to a drastic shift in the temperature of the AFD transition either upward or downward. The latter result puts strong constraints on any hypothetical role played by the AFD soft mode in the formation of Cooper pairs and the emergence of superconductivity in SrTiO$_3$.
\end{abstract}
\maketitle

Strontium titanate, of the ABO$_{3}$ family of perovskites, has continuously attracted attention during the last five decades.  As a band insulator, it displays a fascinating variety of fundamental\cite{Zubko,Koreeda} and application-oriented\cite{Mavroides,Rice} properties. It is a quantum paraelectric, avoiding a ferroelectric transition thanks to quantum fluctuations\cite{Muller1979}. With a low-temperature electric permittivity enhanced by four orders of magnitude, it turns to a metal upon the introduction of infinitesimal concentration of dopants\cite{Spinelli} or even after shining light\cite{Kozuka}. The carriers are remarkably mobile\cite{Son}, as a consequence of an exceptionally long Bohr radius\cite{Behnia} and scatter off each other generating a resistivity which varies as $T^2$, with a very large prefactor\cite{Lin2015}. In spite of being far apart, the electrons living in this peculiar environment can form Cooper pairs in the most dilute superconductor currently known\cite{Lin2013}.

After five decades of intensive research, several questions remain unanswered. What causes superconductivity\cite{Gorkov2015,Edge2015,Ruhman2016} and restricts it to a concentration range of $10^{-5}<n< 2 \times10^{-2}e^{-}/f.u.$\cite{Lin2013}? Why is the inelastic scattering so large \cite{Lin2015}? Does the aborted ferroelectricity\cite{Muller1979} play a role in superconductivity because of its quantum criticality\cite{Rowley2014}?

In contrast to its isovalent siblings BaTiO$_{3}$ and CaTiO$_{3}$, SrTiO$_{3}$ is cubic at room temperature. However, it becomes tetragonal below 105 K. In the ordered state, TiO$_{6}$ octahedra rotate in alternating directions around the $c$-axis\cite{Lytle,Unoki,Fleury,Muller1968}. This antiferrodistortive phase transition has been studied by a variety of experimental techniques. Clear anomalies in specific heat\cite{Salje1998}, thermal conductivity\cite{Steigmeier}, sound velocity\cite{Bauerle} and thermal expansion \cite{Tsunekawa} have been observed at $T_\text{AFD}\simeq$105 K. The transition does not break inversion symmetry, but the boundaries between tetragonal domains become polar\cite{Scott,Salje2013} deep inside the ordered state. In spite of being widely documented over several decades, the consequences of this phase transition for mobile electrons have been elusive.  A recent study\cite{Verma2014} concluded that the AFD transition should be accompanied by en enhancement in electron scattering. However, the experimental data does not show any visible anomaly at the transition temperature. This is also the case of other numerous studies devoted to the transport properties of metallic SrTiO$_{3}$ , which failed to detect any anomaly in the vicinity of 105 K in electric\cite{Spinelli,Son,Lin2015,Cain} or thermoelectric\cite{Cain}transport.

\begin{figure}
\resizebox{!}{0.6\textwidth}
{\includegraphics{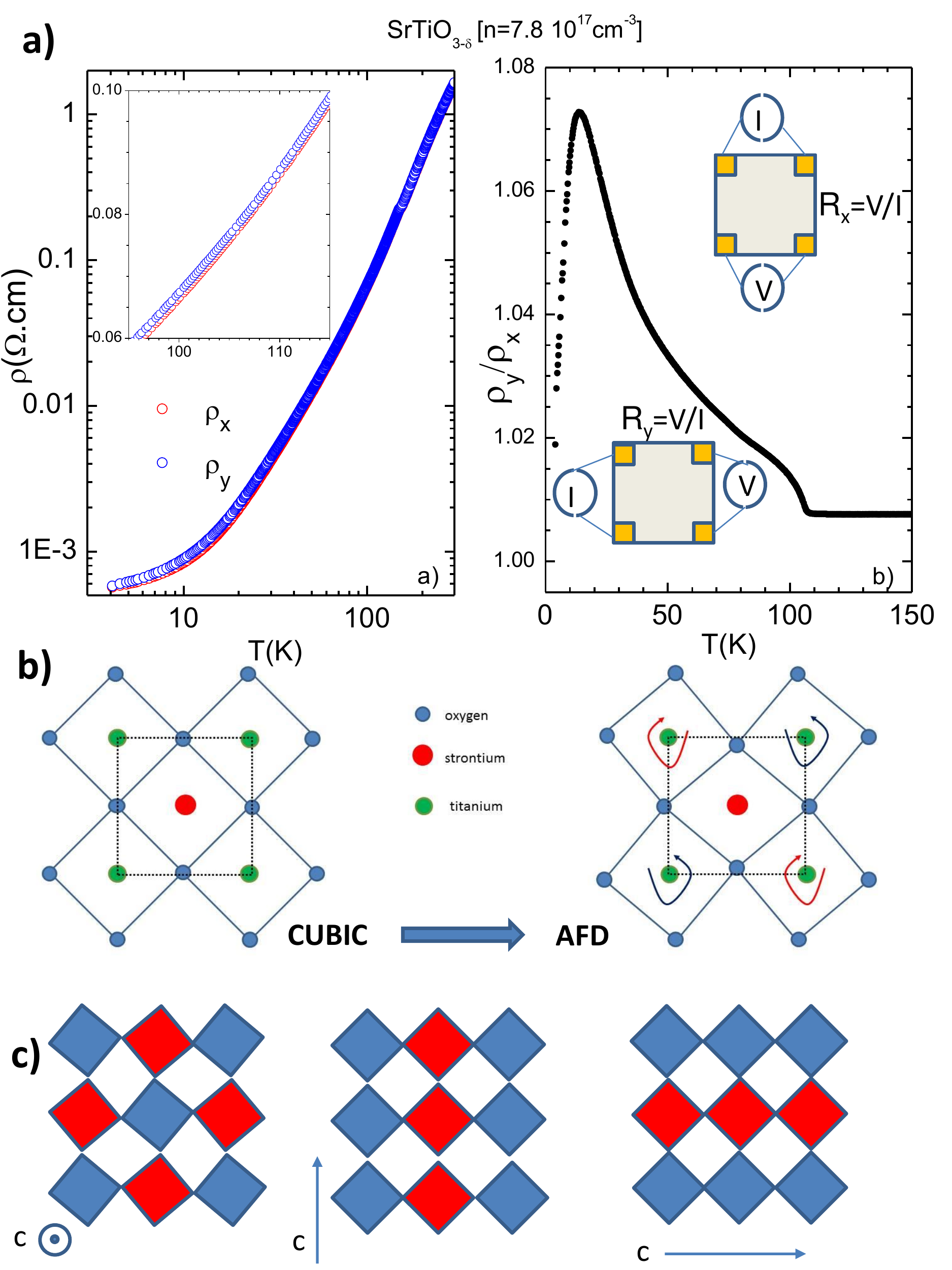}}
\caption{\label{Fig.1}  a) Left: Semi-log plot of $\rho(T)$ in lightly-doped SrTiO$_{3-\delta}$ along two perpendicular axes. The inset is a zoom near the structural transition. No clear anomaly is visible  at 105 K. Right: The $\rho_{y}/\rho_{x}$ ratio shows a clear upward deviation from a constant value, close to unity, at 105 K. The inset shows the geometry of the Montgomery method (see text). b) Atomic configurations in the cubic and antiferrodistortive states. In the latter adjacent octahedra rotate clockwise and anti-clockwise. c) Representations of the three possible domains in the AFD state according to the orientation of the c-axis. Blue(red) squares represent octahedra which rotate clockwise (anti-clockwise).}
\end{figure}

In this paper, we report on a direct measurement of the in-plane anisotropy of charge conductivity in metallic SrTiO$_{3}$ with different dopants. Thanks to the sensitivity of the Montgomery technique\cite{Montgomery} employed here for the first time to study SrTiO$_{3}$, a sharp anomaly at 105 K is clearly visible. The very slight departure from cubicity ($c/a < 1.0006$) gives rise to an anisotropy of conductivity several orders of magnitude larger ($\rho_{c}/\rho_{a}\sim 1.1$). First-principles calculations of the  Fermi surface in the AFD state explain the sign of the anisotropy just below the transition temperature and the order of magnitude of its amplitude. Remarkably, we find that the anisotropy in conductivity displays a non-monotonous temperature dependence. Applying uniaxial strain favors two out of the three possible tetragonal domains and amplifies the non-monotonous behavior. In contrast to what has been observed in other Fermi liquids, elastic and inelastic scattering have opposite sign anisotropies. Finally, we report on the evolution of $T_\text{AFD}$ with doping, and find that different dopants, which give the same superconducting $T_\text{c}$, affect the AFD transition in different ways. We argue that this provides evidence against a theoretical scenario tracing the microscopic origin of superconductivity to the AFD soft phonon mode\cite{Appel}.

The Montgomery technique is a powerful method used to extract the anisotropy of conductivity in a solid shaped as a rectangular prism\cite{Montgomery,Dos_santos}. It is particularly well adapted to the case of SrTiO$_{3}$ because of the availability of relatively large single crystals with precise geometry, on whose corners we evaporate gold pads. At each temperature, current is applied and voltage measured using two pairs of adjacent electrodes in two perpendicular configurations (See the inset of Fig.~1a). The ratio of apparent resistance values ($R_{i}$ = V/I) yields a parameter $x=\ln({R_{2}} / {R_{1}}) / 2\pi$,  which is linked to the ratio of in-plane dimensions ($L_{i}$) of an equivalent isotropic conductor (${L_{2}} / {L_{1}} \cong \sqrt{1+x^{2}}+x$)\cite{Dos_santos}. Assuming that the sample is a perfect square with thickness $e$, the ratio ${L_{2}} / {L_{1}}$ yields the intrinsic resistivity along each perpendicular axis\cite{Dos_santos}:
\begin{equation}\label{2}
 \rho_{1}= \frac{\pi}{8}e \frac{L_{1}}{L_{2}}R_{1}\sinh (\pi\frac{L_{2}}{L_{1}});
 \rho_{2}= \frac{\pi}{8}e \frac{L_{2}}{L_{1}} R_{1}\sinh (\pi\frac{L_{2}}{L_{1}})
\end{equation}

\begin{figure}
\resizebox{!}{0.6\textwidth}
{\includegraphics{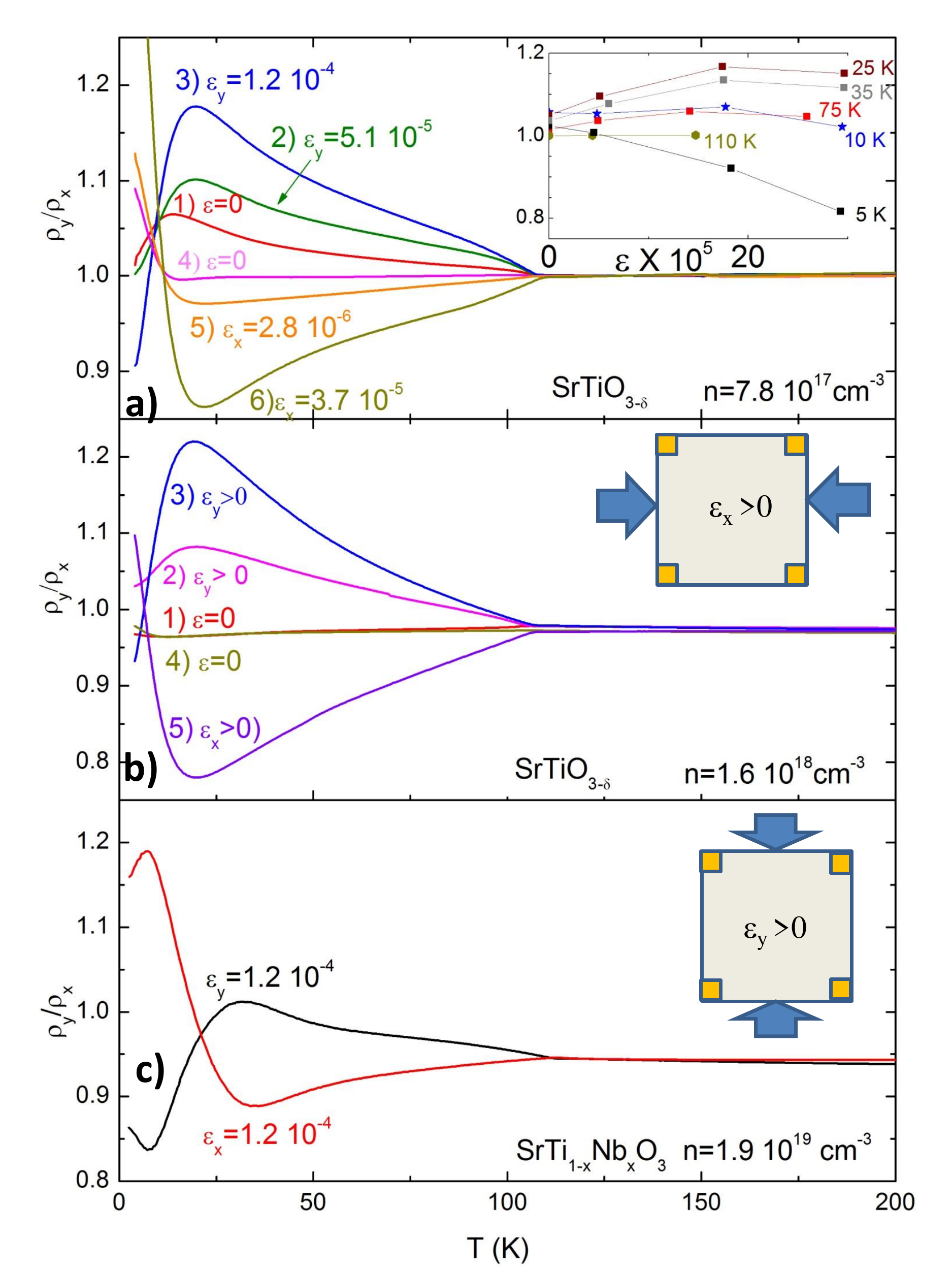}}
\caption{\label{Fig.2} a) and b) The ${\rho_{y}} / {\rho_{x}}$ ratio as a function of T in two SrTiO$_{3-\delta}$ samples with different carrier densities under uniaxial strain. By applying strain along perpendicular orientations, the anisotropy is inverted. Curve indices correspond to measurement sequence and the magnitude and orientation of strain.  The $\epsilon=0$ data corresponds to the strain-free configuration in the beginning and the end of one strain cycle. c) Same for a Nb-doped sample. The inset in a) shows the variation of the anisotropy as a function of strain at different temperatures.}
\end{figure}

Like a previous report\cite{Dos_santos}, we found that the magnitude of resistivity obtained by this method is in agreement with what is obtained by a standard four-contact set-up. The advantage of the Montgomery technique is the direct resolution of the anisotropy, as shown in panels a and b of Fig.~\ref{Fig.1}, which the date for a lightly-doped SrTiO$_{3-\delta}$  (n=7.8 10$^{17}$cm$^{-3}$)sample. It is hard to see any anomaly in the vicinity of 105 K in either $\rho_{x}$ and $\rho_{y}$, but the temperature dependence of their ratio reveals a clear abrupt jump at the onset of the AFD transition.  The ratio is close to unity and temperature-independent above $T_\text{AFD}\simeq$ 105 K and suddenly jumps at $T_\text{AFD}$. The slight departure from unity above $T_\text{AFD}$ presumably results from imperfect squareness of the sample, the finite size of the contact size and other technical limitations. To the best of our knowledge, this is the first detection of any consequence of the AFD transition for charge transport in strontium titanate. The Montgomery technique appears as a powerful method to detect the loss of rotational symmetry and the emergence of nematicity \cite{Fradkin} in other solids such as iron-based superconductors\cite{Kuo}.

Remarkably, multiplicity of domains does not hinder the observation of the symmetry breaking by a directional probe such as resistivity. Each of the three axes of the elementary cubic cell can become the tetragonal axis (Fig.~ 1c). Therefore, there are three possible domains in an unconstrained crystal and the $\frac{\rho_{y}}{\rho_{x}}$ ratio is not the intrinsic anisotropy of SrTiO$_{3-\delta}$. Its magnitude is set by the accidental imbalance between the population of these domains. One would expect therefore that the magnitude and the orientation of applied strain should modify the $\frac{\rho_{y}}{\rho_{x}}$ ratio; as seen in Fig. 2, this is indeed the case. Cooling the same sample in the presence of  strain, the anisotropy in the ordered state is enhanced. Moreover, rotating the orientation of the applied strain by 90 degrees in the same sample yields an opposite sign for ($\frac{\rho_{y}}{\rho_{x}}-1$) in the AFD state. We conclude that the samples are multi-domain and applying compressive strain favors the two domains which have their (longer) $c$-axis perpendicular to the strain. As seen in Fig. 2c, strained SrTi$_{1-x}$Nb$_{x}$O$_{3}$ displays also a sizeable anisotropy in its $\frac{\rho_{y}}{\rho_{x}}$ ratio below the transition temperature.

The first outcome of this study is a quantification of the magnitude of $\frac{\rho_{y}}{\rho_{x}}$ ratio. As seen in Fig. 2, in samples subject to compressive strain, this ratio can become as large as 1.2 . At first sight, it looks surprising that an anisotropy of this amplitude in charge conductivity results from a tetragonal distortion quantified by x-ray diffraction to be as small as 1.00056\cite{Lytle}. The tetragonal distortion is linked to the angle of rotation of the TiO$_{6}$ octahedra\cite{Unoki} :

\begin{equation}\label{3}
\frac{c}{a}=\frac{1}{\cos \varphi}
\end{equation}

The measured values of $\varphi=2.1^\circ$\cite{Unoki} and $c/a$=1.00056 \cite{Lytle} satisfy this expression. As seen in  Fig. 1c, the slight tetragonal distortion is accompanied by a more substantial misalignment between Ti atoms. Considering that their t$_{2g}$ orbitals host the threefold degenerate conduction bands of metallic SrTiO$_{3}$, the consequences can be drastic.  A slight misalignment between adjacent TiO$_{6}$ octahedra can drastically change the overlap between T$_{2g}$ orbitals of two neighboring titanium atoms. Thus, a very small $\varphi$ can cause a substantial drop in conductivity along the misalignment orientation providing a qualitative agreement for the experimental observation. As we will see below, according to first-principles band calculations, the Fermi surface topology is deeply affected by the phase transition.

A second and more unexpected result is the temperature dependence of the anisotropy in the ordered state. As seen in figures  \ref{Fig.1} and \ref{Fig.2}, the temperature dependence of $\frac{\rho_{y}}{\rho_{x}}$ ratio is not monotonous. In SrTiO$_{3-\delta}$, it peaks around 20 K and then decreases to below unity. In SrTi$_{1-x}$Nb$_{x}$O$_{3}$, the temperature dependence is even more complex, with a peak and a shallow minimum. If the $\frac{\rho_{y}}{\rho_{x}}$ ratio were proportional to an order parameter, then this would suggest a second phase transition well below 105 K. The existence of additional phase transitions in SrTiO$_{3}$ has been a longstanding controversy. However,  our observation can be accounted for if carrier scattering by phonons, electrons and defects do not share the same anisotropy. We will come back to this below.

\begin{figure}
\resizebox{!}{0.375\textwidth}
{\includegraphics{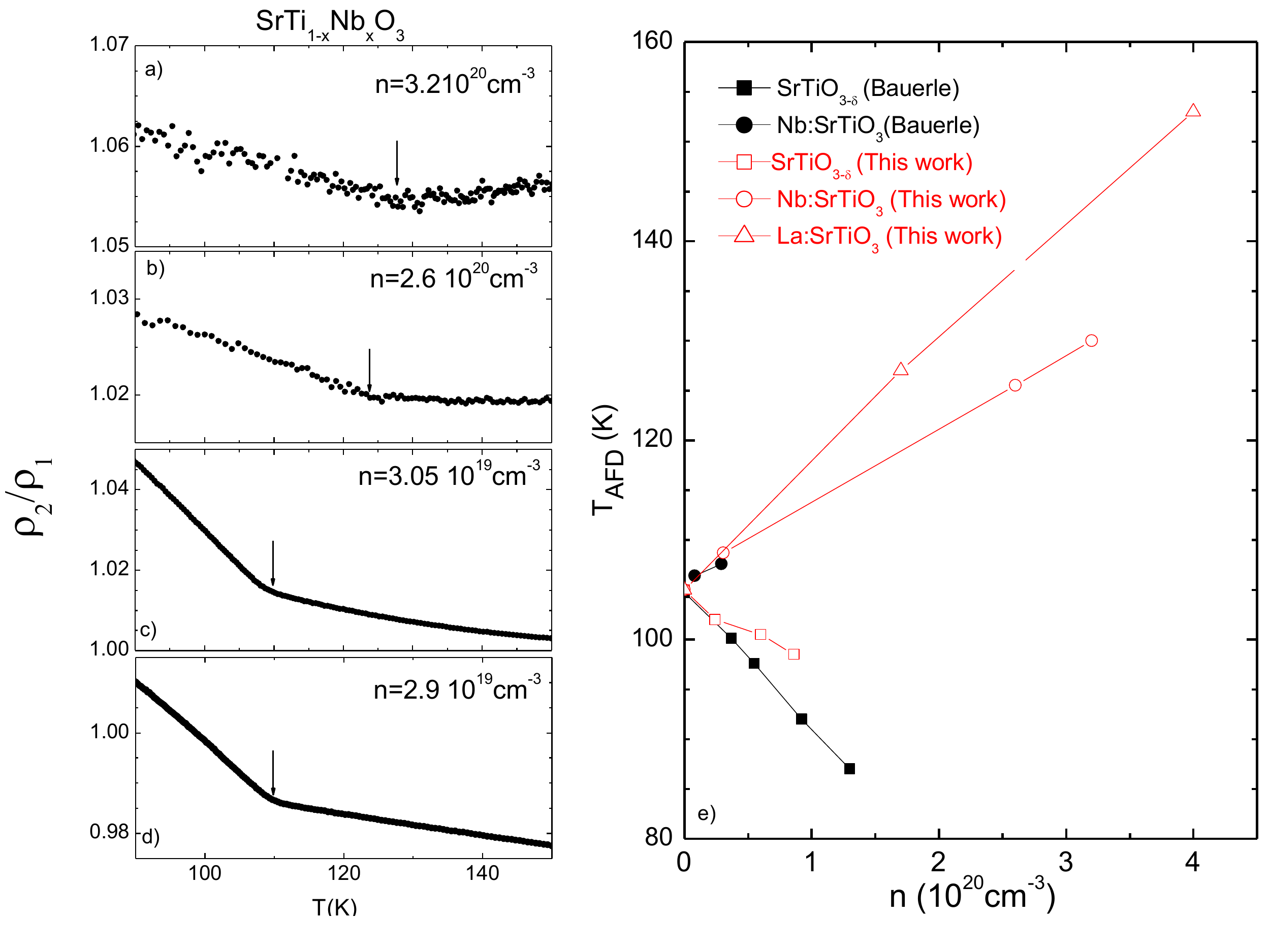}}
\caption{\label{Fig.3} a) to d): The evolution of the resistivity anisotropy anomaly in temperature for SrTi$_{1-x}$Nb$_x$O$_3$. The anomaly shifts to higher temperatures with increasing doping. e) Evolution of the AFD transition temperature with carrier concentration. The data reported in Ref.\cite{Bauerle} is also included for comparison. Opposite trends are observed for alternative routes for doping. While Nb (or La) substitution enhances $T_\text{AFD}$, the introduction of oxygen vacancies depresses $T_\text{AFD}$.}
\end{figure}

We now turn our attention to the evolution of the AFD transition temperature with doping. As seen in Fig.~3, the AFD anomaly in $\frac{\rho_{y}}{\rho_{x}}$ ratio shifts to higher temperature with the substitution of Ti with Nb. We also studied  the evolution of $T_\text{AFD}$ with oxygen reduction and La substitution. Our results are in agreement with previous studies of the AFD transition using sound velocity\cite{Bauerle} and Ramman scattering\cite{Tkachi}.The emerging picture is summarized in Fig. 3e. While substituting Ti with Nb (or Sr with La) leads to an upward shift of $T_\text{AFD}$, introducing oxygen vacancies pulls it downward. The extreme sensitivity of the AFD transition to the presence of a tiny concentration of foreign atoms is remarkable. Substituting Sr with larger La or Ti with larger Nb both strengthen the AFD instability, presumably by weakening the stability of the cubic structure, which relies on a fragile balance of the three atomic radii in SrTiO$_{3}$. Intriguingly, this also happens when Sr is replaced with smaller Ca\cite{Delima}. In contrast, oxygen vacancies weaken the tetragonal distortion, presumably because they destroy the antiferrodistortive coupling between adjacent octahedra.

The contrast in response of $T_\text{AFD}$ to oxygen reduction compared to La and Nb substitution has an unnoticed implication. All three routes for $n$-doping, i.e., oxygen removal, substituting Ti(by Nb)or Sr (by La)\cite{Koonce,Suzuki} lead to a dilute metal with a superconducting $T_\text{c}$ which peaks at a carrier density in the range of $n=10^{20}$ cm$^{-3}$. Decades ago, Appel\cite{Appel} suggested a link between the formation of Cooper pairs and the soft phonon mode associated with the octahedra rotation and the AFD transition\cite{Appel}. The drastic difference in the dependency of  $T_\text{AFD}$ on dopants type indicates that either the soft mode does not play a major role in superconductivity or its response to reduction and substitution is not the same as $T_\text{AFD}$.

\begin{figure}
\resizebox{!}{0.35\textwidth}
{\includegraphics{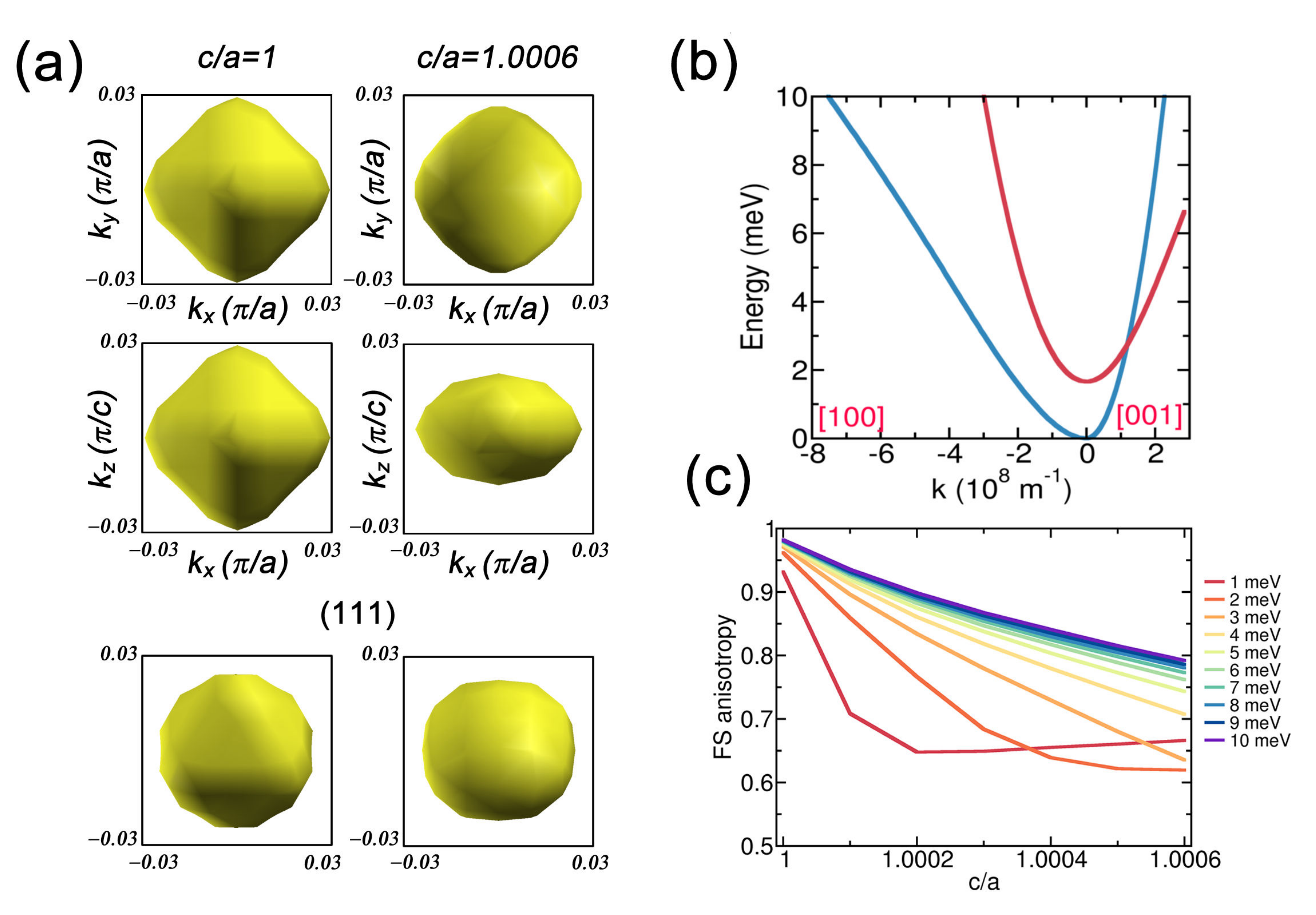}}
\caption{\label{Fig.4} a) Theoretical Fermi surface in the cubic and the tetragonal states ($c/a$=1.0006). $E_\text{F}$ is placed 1 meV above the Conduction Band Minimum. The Fermi surface is viewed along [001], [010], and [111]. b) Band dispersion of STO for $c/a$=1.0006.  c) Fermi surface anisotropy (defined as the ratio of its radii along two perpendicular orientation) as a function of distortion, for different chemical potentials.}
\end{figure}

We carried out \emph{first-principles} calculations to quantify the Fermi surface (FS) anisotropy. Fig.~\ref{Fig.4}  shows the FS cross sections at $E_\text{F}=$1 meV, the electron band structure, and the anisotropy of the FS as a function of $c/a$ for chemical potentials ($\mu$) from 1 to 10 meV. The lower conduction bands in STO originate from the same orbital manifold. In the tetragonal system they have distinct irreducible representations for the spatial degrees of freedom, which allows them to cross, only at a single point along 001. In the plane (or in any skew direction) the AFD rotation mixes their representations, and the bands will anticross, as is seen along 100. In agreement with previous theoretical studies\cite{Vandermarel}, we find that the cubic phase FS has six lobes and in the tetragonal state, two out of these six are squeezed.

%
According to Shubnikov-de Haas (SdH) experiments \cite{Lin2013,Lin2014,Allen} when carrier density is $n \simeq $7.8$\times$10$^{17}$cm$^{-3}$, the Fermi energy is close to $\mu \simeq 1$ meV. At this level (Fig. 4a), we find a FS area in the (001) plane about 1.7 times larger than in the (010) plane, in fair agreement with the SdH data\cite{Allen}. The calculated band masses are lighter than the experimental values, indicating the importance of renormalization by electron-phonon interactions and electron correlation\cite{king2014, vanmechelen2008}. The estimated renormalizations are 1.5 for the lower band (similar to explicit calculations in Ref.~\cite{king2014}) and 3.5 for the upper band. A quantitative description of Fermi surface would requires the calculation of \emph{anisotropic} mass renormalization.  The FS anisotropy, (ratio of its radii along $c$ and $a$) reaches 0.65 (Fig.~4c) for $\mu = 1$ meV. The Fermi radius is shorter (and $v_F$ larger) along $c$, such that the conductivity should be larger as well, in agreement with the experimental observation in the vicinity of $T_\text{AFD}$. Finally, according to the calculations,  anisotropy decreases with increasing $\mu$(see the supplement). This matches the experimental observation of diminishing anomaly at $T_\text{AFD}$ with increasing doping (Fig.~\ref{Fig.3}a-d).

\begin{figure}
\resizebox{!}{0.58\textwidth}
{\includegraphics{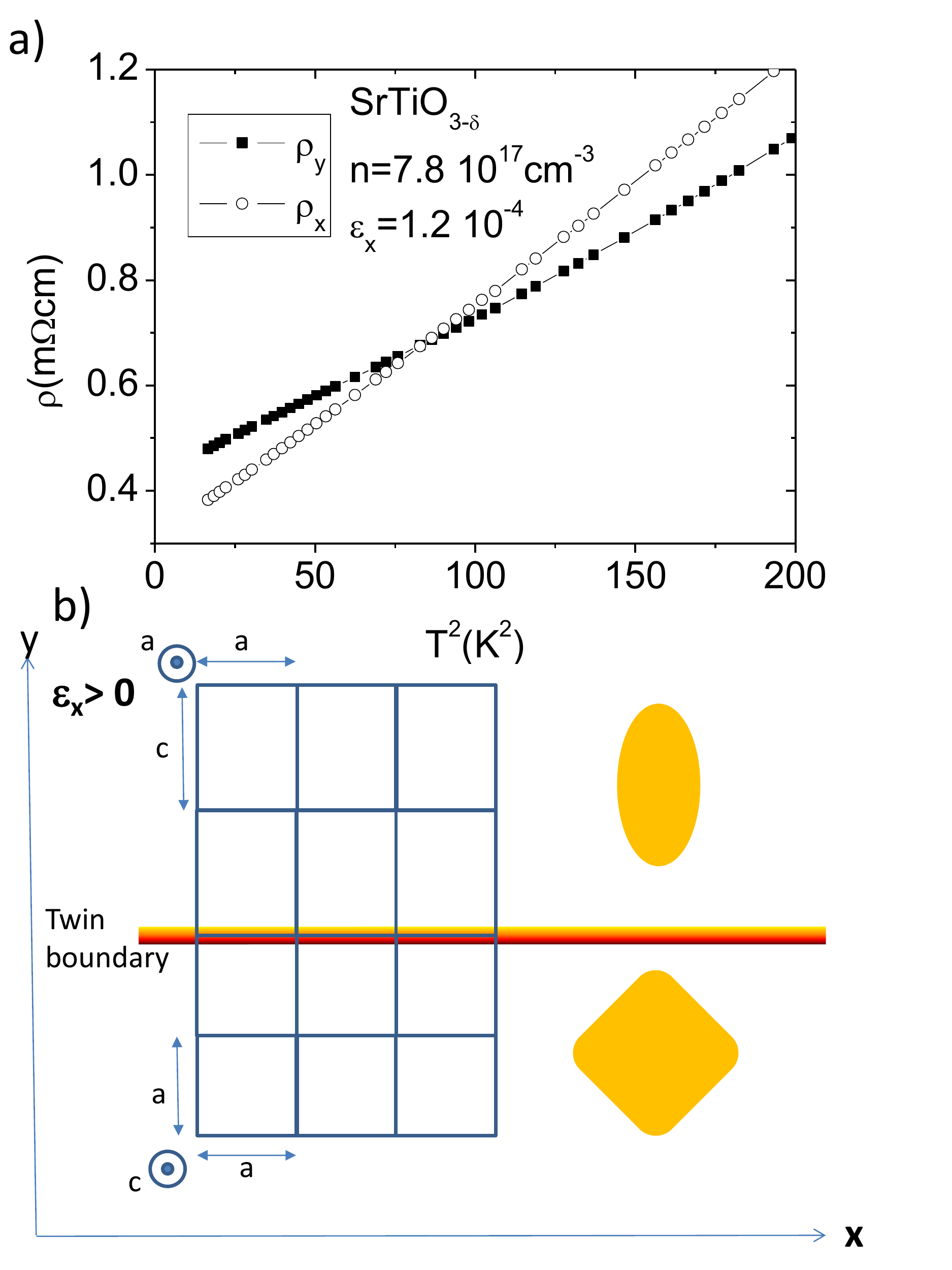}}
\caption{\label{Fig.5} a) $\rho_{y}$  and $\rho_{x}$ as a function of $T^2$ in strained SrTiO$_{3-\delta}$. The anisotropy of the slope and the intercept are opposite, which points to a different anisotropy for elastic and inelastic components of resistivity. b) Applying strain along $x$ favors the two strain domains whose longer $c$-axis is perpendicular to $x$. In the (x, y) plane, the twin boundary between these two domains is parallel to $x$ and scatters less electrons traveling along $x$, leading to a higher residual resistivity. On the other hand, the average Fermi velocity, set by the anisotropy of the Fermi surface is larger along $x$. This leads to a lower inelastic resistivity  for this orientation. }
\end{figure}

Let us now examine the puzzle of the non-monotonic temperature dependence of $\rho_y/\rho_x$. As seen in Fig. 1a, upon cooling from room temperature to liquid-helium temperature, the resistivity of metallic SrTiO$_{3}$ drops by three orders of magnitude. Thus, inelastic scattering of electrons is by far the dominant scattering around 100 K. Most of this inelastic scattering is due to phonons, but below 30 K, the electron-electron scattering leads to a quadratic temperature dependence. Fig. 5 shows the low-temperature resistivity of a strained oxygen-reduced sample as a function of $T^{2}$. In agreement with previous reports\cite{Vandermarel, Lin2015,Mikheev}, $\rho_{x}$ and $\rho_{y}$ both follow a behavior described by $\rho_{i}= \rho_{0i}+A_{i}T^2$ (i=$x$,$y$). The anisotropy of the prefactor is set by the expected anisotropy of conductivity stemming from the topology of the Fermi surface. Note that in this temperature window, below the degeneracy temperature of electrons, the relevance of the Fermi surface topology is straightforward. On the other hand, when the carrier density is as low as $ n=7.8 \times 10^{17}cm^{-3}$, the system is non-degenerate and subject to Boltzmann statistics at 105 K. But, even in this non-degenerate regime, the anisotropy of conductivity still keeps the trace of the anisotropy of the Fermi velocity. The prefactor of T-square resistivity along a-axis (c-axis) is  $A_{a}=4.9\mu\Omega cmK^{-2}$($A_{c}=3.4\mu\Omega cmK^{-2}$) [The average ($A_{av}= (2A_{a}+A_{c})/3= 4.4\mu\Omega cmK^{-2}$) is close to what was measured  for a sample of comparable carrier density by a conventional four-probe technique\cite{Lin2015}]. As expected for an  anisotropic Fermi liquid, for example in UPt$_{3}$\cite{Joynt}, the ratio ($A_{a}/A_{c}\simeq 1.5 $) is close to the ratio of the Fermi velocities along $c$-axis and $a$-axis.

As seen in Fig. 5, the intercept (i.e. the residual resistivity, $\rho_{0}$) is lower along the direction where the prefactor $A$ is larger. In other words, elastic scattering is larger along the orientation for which inelastic scattering is smaller. This generates a peak in $\frac{\rho_{y}}{\rho_{x}}$($T$), since the balance between the two components of the resistivity is changed at an intermediate temperature. This feature distinguishes SrTiO$_{3}$ from other anisotropic Fermi liquids. In the latter cases, the residual resistivity and the prefactor of the $T^{2}$ resistivity share the same anisotropy, set by the Fermi velocity anisotropy.  The same is true of UPt$_{3}$\cite{Joynt}as well as two other well-known perovskites, the ruthenate  Sr$_{2}$RuO$_{4}$\cite{Mackenzie} and cuprate La$_{1.7}$Sr$_{0.3}$CuO$_{4}$\cite{Nakamae}. In both of these layered conductors, the prefactor of the quadratic term in the resistivity and the residual resistivity share the same anisotropy.

The exceptional case of metallic SrTiO$_{3-\delta}$ may be a result of the particular orientation of boundaries between twins in a sample subject to uniaxial strain (see Fig. 5b). Application of strain along one direction leads to the suppression of one domain out of three (the one with its $c$-axis along the strain). Since the Fermi velocity is larger along the $c$-axis, the conductivity should become \emph{lower} in the orientation parallel to the applied strain.  This is indeed what is seen experimentally. On the other hand, in such a configuration the favored domain boundary would lie along the orientation of compressive strain. As a scatterer, such a boundary would introduce additional scattering perpendicular to the direction of strain. This could explain the opposite signs of $(\rho_y-\rho_x)/\rho_x$ at the onset of the transition (governed by the anisotropy of the Fermi velocity) and in the zero temperature limit (set by the orientation of twin boundaries).

Two other features should also be included in an exhaustive quantitative treatment of the problem. The first is the experimental evidence for enhanced conductivity along a twin boundary in SrTiO$_{3}$\cite{Kalisky}, amplifying the anisotropy introduced by the presence of a twin boundary along the orientation of compressive strain. The second is the contrast between potential wells introduced by Nb dopants and oxygen vacancies. They are expected to  differ in structure and the angular distribution of the scattering they cause.

Several other questions remain. The precise shape of the Fermi surface in dilute metallic SrTiO$_{3}$ is not settled yet. A detailed angle-dependent study on single-domain samples is necessary to map the full FS. The quantification of the non-parabolic dispersion is important to pin down the origin of the $T^2$ resistivity in the dilute limit\cite{Lin2015}. Non-parabolicity leads to a distinction between momentum and velocity, and breaks Galilean invariance\cite{Pak}, which is expected to play a significant role in the $T^{2}$ resistivity of SrTiO$_{3}$.

In summary, we quantify the anisotropy of charge conductivity caused by the AFD transition in metallic SrTiO$_{3}$, using the Montgomery technique. Both first-principles band calculations and experimental data find that the very slight departure from cubicity gives rise to a sizeable anisotropy in the Fermi surface and conductivity. In contrast with other anisotropic Fermi liquids, elastic and inelastic resistivity do not present the same anisotropy. The contrasting response of the AFD and superconducting phase transitions to the type of dopant argues against any role played by the octahedral rotational mode in superconductivity.

This work was supported by Agence National de la Recherche through SUPERFIELD and QUANTUMLIMIT projects. Q.T. acknowledges a grant offered by China Scholarship Council and support by the National Basic Research Program of China (Grant No.2012CB821404). B. X. and M. J. V. are supported by ARC projects TheMoTherm (GA 10/15-03) and AIMED (GA 15/19-09) from the Communaut\'e fran\c{c}aise de Belgique, and a PDR project (GA T.1077.15) from the Fonds National pour la Recherche Scientifique (Belgium). Computer time was provided by CECI (FNRS GA 2.5020.11), SEGI, and CENAERO/ZENOBE through the Walloon Region GA 1117545.

\appendix
\section{Experimental details}
Oxygen vacancies were introduced in commercial single crystals of SrTiO$_{3}$ by heating them in the temperature range of 800 to 1100 degrees Celsius in a vacuum in the range of 10$^{-7}$ mbar in a manner reported before\cite{Spinelli}. The samples' typical dimensions were  $10 \times 10 \times 0.5 mm ^3$ or $5 \times 5 \times 0.5 mm ^3$. Ohmic contacts on sample corners were made by evaporating gold and heating up to 600 degrees. Compressive uniaxial stress was applied by sandwiching the sample between two metallic blocks connected with two stainless steel screws. The magnitude of applied strain was measured using a 350$\Omega$ resistive strain gage (Kyowa KFL-1-350-C1-11).

\section{Computational details}

A 20-atom cell ($\sqrt{2}\times\sqrt{2}\times2$) is adopted to calculate the Fermi surfaces and band structures of the tetragonal AFD phase of SrTiO$_3$, while a 5-atom unit cell is used for the cubic phase. To study the effect of $c/a$ ratio, the in-plane lattice constant is fixed to an experimental value at 50 K, i.e., 5.507 \cite{Jauch}, while $c/a$ varies from 1 to 1.0006. The AFD tilting angle for each $c/a$ is fixed according to Eq. 2 of the main article. The calculations are carried out within the framework of density function theory (DFT) in the local density approximation (LDA), using the \textsc{abinit} package\cite{Gonze1,Gonze2}. The wave functions are expanded using plane-wave basis sets with kinetic energy cutoff of 45 Hartree. The self-consistent calculations are performed on an unshifted 6$\times$6$\times$4 and 12$\times$12$\times$12 $k$-point grid for the tetragonal and cubic phases, respectively.

The accurate Fermi surface anisotropy (Fig. 4c of the main text) is determined from the band structures along the pseudo-cubic [100] and [001] directions (Fig. 4b of the Article), by finding $k$ for which the band intersects the chemical potential. Optimized norm conserving Vanderbilt pseudopotentials are used here\cite{Hamann}, in order to include the spin-orbit coupling (SOC), which yields a band edge splitting (due to tetragonal distortion) that is in good agreement with experiment \cite{Allen}. This agreement is somewhat fortuitous, given the neglect of electron-electron interactions beyond semilocal DFT and of electron-phonon interaction effects, as already observed by Allen\cite{Allen} and van Mechelen\cite{vanmechelen2008}. Both effects will increase the effective mass, and may contribute to shift the AFD splitting. Larger values of splitting were found with LAPW\cite{Vandermarel}, which should be more accurate DFT as it does not pseudize the wave functions. Using the experimental values for lattice constant and AFD angles removes the additional effect of DFT structural relaxation. The Fermi surface shape (Fig. 4a) is obtained with explicit calculations including 200$\times$200$\times$200 $k$ points in a restricted cube of $\pm 0.05 \pi / a$ in each direction around $\Gamma$.

\begin{figure*}\centering
\resizebox{!}{0.65\textwidth}
{\includegraphics{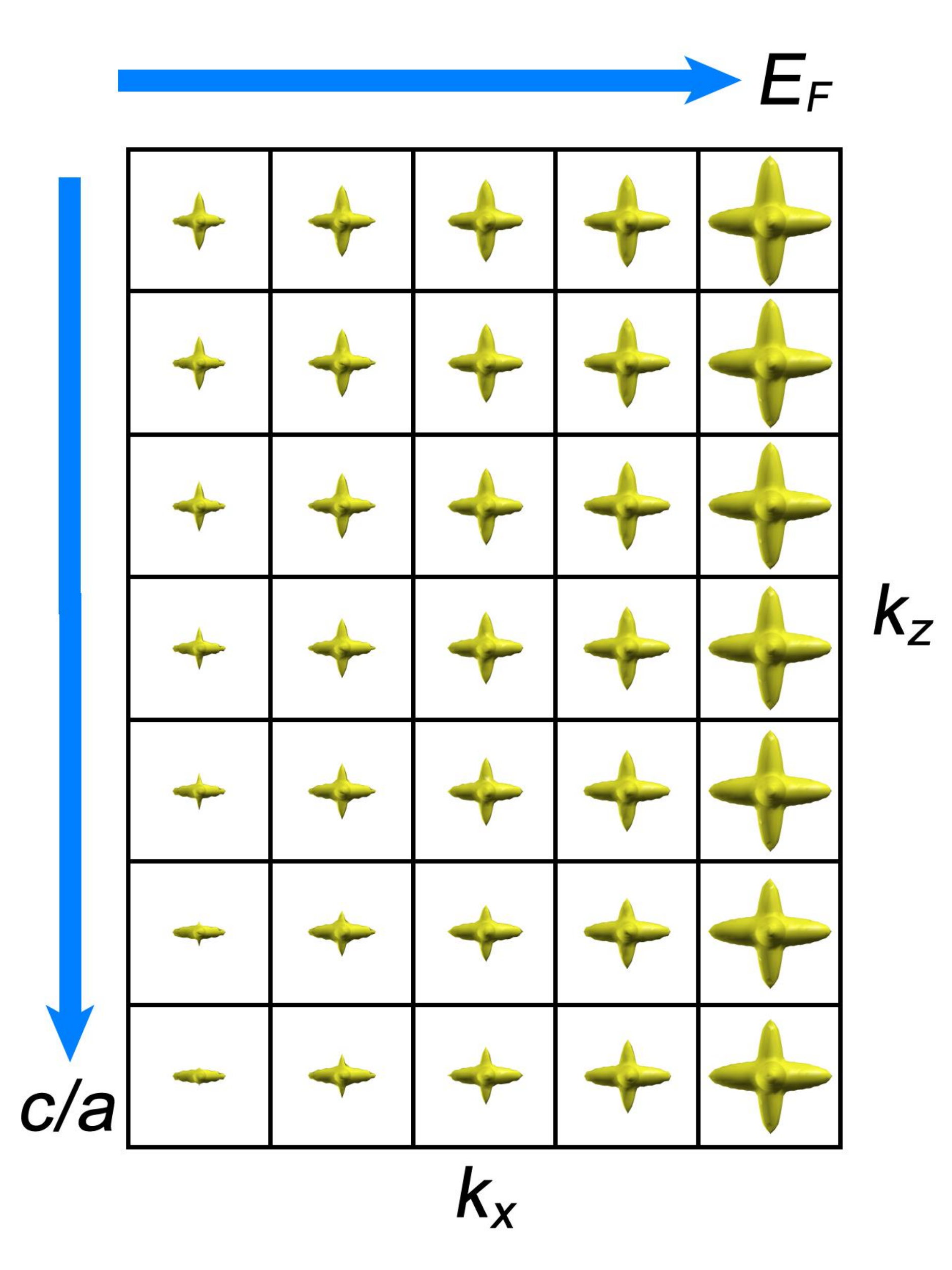}}
\caption{\label{Fig.S1} \textbf{Theoretical Fermi surface anisotropy.} Evolution of the Fermi surface (in the $a$-$c$ plane) with lattice distortion, $c/a$, and chemical potential, E$_{F}$. The distortion varies from 1 to 1.0006  and the values for chemical potential are 2, 3, 4, 5, and 10 meV.}
\end{figure*}

For the graphical representation of the Fermi surface evolution as a function of the chemical potential and AFD angle, non-relativistic extended norm conserving pseudopotentials are adopted \cite{Teter}. The Fermi surfaces of the lowest conduction band are obtained with 200$\times$200$\times$200 $k$ grid in the full Brillouin Zone, using maximally-localised Wannier functions and the \textsc{Wannier90} code\cite{Mostofi}. Without the SOC the FS are not quantitative, but qualitatively show the evolution of the shape and anisotropy.
In agreement with previous theoretical studies\cite{Vandermarel}, we find that the cubic phase FS has six lobes at moderate to high doping. In the tetragonal state, two out of these six are squeezed (Fig. \ref{Fig.S1}). When the chemical potential ($\mu$) is near the band edge, the FS shape is in agreement with the model bands given in \cite{vanmechelen2008}.

\end{document}